# Effect of rare-earth (Er and Gd) substitution on the magnetic and multiferroic properties of $DyFe_{0.5}Cr_{0.5}O_3$

Mohit K. Sharma[1], Tathamay Basu[2, #], K. Mukherjee[1,*] and E. V. Sampathkumaran[2]

[1]School of Basic Sciences, Indian Institute of Technology Mandi, Mandi 175005, Himachal Pradesh, India

[2]Tata Institute of Fundamental Research, Homi Bhabha Road, Colaba, Mumbai 400005, India

*Email: kaustav@iitmandi.ac.in

## ABSTRACT

We report the results of our investigations on the influence of partial substitution of Er and Gd for Dy on the magnetic and magnetoelectric properties of $DyFe_{0.5}Cr_{0.5}O_3$, which is known to be a multiferroic system. Magnetic susceptibility and heat capacity data, apart from confirming the occurrence of magnetic transitions at ∼ 121 and 13 K in $DyFe_{0.5}Cr_{0.5}O_3$, bring out that the lower transition temperature only is suppressed by rare-earth substitution. Multiferroic behavior is found to persist in $Dy_{0.4}Ln_{0.6}Fe_{0.5}Cr_{0.5}O_3$ (Ln= Er and Gd). There is an evidence for magnetoelectric coupling in all these materials with qualitative differences in its behavior as the temperature is changed across these two transitions. Remnant electric polarization is observed for all the compounds. The most notable observation is that electric polarization is seen to get enhanced as a result of rare-earth substitution with respect to that in $DyFe_{0.5}Cr_{0.5}O_3$. Interestingly, similar trend is seen in magnetocaloric effect, consistent with the existence of magnetoelectric coupling. The results thus provide evidence for the tuning of magnetoelectric coupling by rare-earth substitution in this family of oxides.

**INTRODUCTION**

Multiferroic materials are characterized by two or more long-range ordering phenomena e.g., magnetic, electric and/or elastic properties. Such materials have been the subject of extensive research in recent years as these provide a platform to study the interesting physics associated with the mutual coupling between spin and lattice degrees of freedom and due to potential technological applications [1, 2]. In this context, various transition metal oxides like orthomanganites [3], orthoferrites [4], nickelates and cobaltites [5], doped defossities [6], vanadites [7] etc which show multiferroic properties have been investigated. Though rare-earth orthochromites (LnCrO$_3$, Ln = rare-earths) have been studied for several decades for their interesting physical properties [8-13], the interest to search for multiferroic behaviour and magnetoelectric (ME) coupling occurred during last few years only. Such investigations revealed multiferroic behaviour in these materials [14-15]. A new class of half-doped orthochromites, LnFe$_{0.5}$Cr$_{0.5}$O$_3$, Ln = rare-earths [16], have also been studied because of potential application in magnetic data storage, multiple-state memories and magnetic refrigeration [17-21].

In this article, we focus on half-doped orthochromite, DyFe$_{0.5}$Cr$_{0.5}$O$_3$, which was established to be a multiferroic material, exhibiting an enhanced magnetocaloric effect due to magnetoelectric coupling [19]. However, reports about the evolution of magnetic and multiferroic properties due to the nature of rare-earth ions in such a half-doped orthochromite is lacking in literature. Therefore, we have investigated partially doped materials of this oxide (at the Dy site by Gd and Er) to explore how the structural, electric and magnetic properties are affected. DyFe$_{0.5}$Cr$_{0.5}$O$_3$ is found to undergo magnetic transitions at ~ 261, 121 and 13 K, as reported in Ref. [19, 20]. Here, we focus on the magnetic transitions around 13 and 121 K only, as the effect of rare-earth substitution on the magnetic, multiferroic and magnetocaloric properties cannot be clearly tracked due to the weakness of the magnetic features around the 261 K magnetic transition. It is observed that the magnetic transition-temperature of 121 K is unaffected, but the lower transition-temperature is suppressed due to partial replacement of Dy by Er/Gd. Multiferroic behavior persists in the Er and Gd doped compounds as well. Dielectric anomalies are seen at the magnetic ordering temperatures in all these oxides. The coupling strength is seen to get enhanced as a result of rare-earth doping. Both magnetocaloric effect and electric polarization ($P$) are also seen to increase as a result of rare-earth substitution, thereby establishing the role of rare-earth doping on ME coupling.

**EXPERIMENTAL DETAILS**

The polycrystalline samples, DyFe$_{0.5}$Cr$_{0.5}$O$_3$ (DFCO), Dy$_{0.4}$Er$_{0.6}$Fe$_{0.5}$Cr$_{0.5}$O$_3$ (DEFCO) and Dy$_{0.4}$Gd$_{0.6}$Fe$_{0.5}$Cr$_{0.5}$O$_3$ (DGFCO) were prepared by solid state reaction of stoichiometric amounts of Dy$_2$O$_3$, Er$_2$O$_3$, Gd$_2$O$_3$, Fe$_2$O$_3$ and Cr$_2$O$_3$ (Sigma-Aldrich, purity > 99.9%), with the same synthetic conditions as described in Ref. [21]. The structural analysis was carried out by x-ray Diffraction (XRD, Cu K$_\alpha$) using Rigaku Smart Lab instruments. The Rietveld refinement of the powder diffraction data was performed by FullProf Suite software [22]. Temperature (2-300 K) and magnetic field (up to $H$= 50 kOe) dependent magnetization ($M$) was measured with a Magnetic Property Measurements System (MPMS), Quantum design, USA. Heat capacity ($C$) measurements in the temperature range 2-150 K were performed using the Physical Property Measurements System (MPMS). Complex dielectric permittivity was measured as a function of $T$ (2-150 K) and $H$ using Agilent E-4980A LCR meter with a homemade sample holder integrated with PPMS. The ac bias was 1 V and different frequencies ($\upsilon$= 30 to 500 kHz) were used. The data was collected in the warming cycle (at the rate of 1 K/min). Electric polarisation was determined from the pyroelectric current measured with Keithley 6517A electrometer.

**RESULTS AND DISCUSSION**

Figure 1(a-c) shows the XRD patterns of the samples under investigation. A comparison of the XRD pattern of the parent and doped compounds (inset of Fig. 1 (b)) reveals a shift of diffraction lines to the lower-angle and higher-angle side with Gd and Er doping respectively, following lanthanide contraction. This establishes that the dopants go to respective sites. Rietveld refinement (see the lines through the data points) confirmed formation of the samples in the single phase with orthorhombic perovskite structure (Pbnm space group). Table 1 summarizes the relevant structural parameters obtained by fitting the powder XRD data by Rietveld refinement. The lattice parameter of DFCO matches well with that reported in the literature [19]. The changes in the lattice parameter are ≤0.5% due to the substitution. The low value of $R$- factors and goodness of fit shows good agreement between calculated and observed patterns.

Figure 2(a) shows the temperature dependence of dc magnetic susceptibility ($\chi$=$M/H$) of all the samples in the $T$-range 2 to 300 K measured with 100 Oe for the zero-field-cooled (ZFC) condition of the specimens. The data for DFCO is in agreement with the literature [19, 20] that this compound undergoes distinctly two magnetic transitions, one around ~121 K ($T_1$) and the other at ~13 K ($T_2$). The magnetic feature around 261 K, reported in Ref. 19 is very weak in our samples, and it could be seen in the d$\chi$/dT vs. $T$ plot only (inset of Fig 2(a)). The derivative plots in figure 2 (c) also bring out the transitions around 13 K and around 120 K. In the rest of the article, we now focus on the transition temperatures $T_1$ and $T_2$ only. Interestingly, for both DEFCO and DGFCO, $T_1$ remains unchanged and $T_2$ shifts downwards to ~ 9 and 4 K respectively, as inferred from the derivative plots. The magnetic ordering in DFCO at $T_1$ is ascribed to Cr-O-Cr magnetic interactions, whereas the one at $T_2$ is due to Dy-O-Fe/Cr magnetic interactions [19]. Hence it can be concluded that the rare earth substitution does not influence the magnetic ordering due to Cr. There is a marginal, but non-negligible, decrease in $T_2$ for DEFCO, but the observed decrease is more dramatic for Gd substitution. Thus the observed trend in $T_2$ does not follow lanthanide contraction. The depression of $T_2$ can be attributed to the dilution of Dy sublattice, possibly, due to the magnetic moment of low-lying crystal-field ground state of the dopant, as Er is expected to undergo crystal-field splitting unlike Gd in the temperature range of interest. Alternatively, Gd and Er could be on the verge of magnetic ordering as the temperature is lowered to 2 K and the internal magnetic field generated due to this, however small it may be, is such that it depresses antiferromagnetic ordering due to Dy.

Isothermal magnetization behavior as a function of $H$ was obtained up to 50 kOe in the temperature range 2 to 150 K. Representative curves at different $T$ (= 2, 40 and 150 K) are shown in Fig. 2(b), (d) and (f). For all the samples, a weak magnetic hysteresis is observed at low fields (≤ 5 kOe) in all these curves and the magnetization does not saturate at high fields. The non-saturation tendency of magnetization is usually noted for systems with finite antiferromagnetic coupling, whereas the presence of magnetic hysteresis is a signature of the presence of ferromagnetic component. For DFCO and DEFCO, at 2 K, the virgin curve shows a change in curvature indicating a weak field-induced metamagnetic effect (the feature is more clearly seen in the d$M$/d$H$ vs $H$ plot (inset of Fig. 2-(e)). With increasing temperature, this feature is suppressed, as seen from the $M$-($H$) plot at 40K. For DGFCO, at 2 K, the $M$-($H$) curve is a smooth S-shaped one without any evidence for metamagnetic effect; at 40 K, $M$ increases linearly with $H$. From the behavior of $M(H)$ curves at 2 K, we infer that the magnetic behavior of half-doped orthochromite is sensitive to the nature of rare-earth ions. This fact is further substantiated by the temperature response of coercive field ($H_C$) (Fig 2(e)).

The nature of variation of $H_C$ with $T$ is found to be different for different compositions. For DFCO, $H_C$ initially decreases up to 20 K and then it increases with increasing $T$. However for DEFCO, $H_C$ increases with increasing temperature up to 40 K, beyond which it decreases. In the case of DGFCO, $H_C$ exhibits a continuous increase up to 120 K and then it shows a decreasing trend. It is possible that relative alignment of transition-metal and rare-earth sub-lattice moments also may be influenced by the variation of $H$ and $T$ [9, 10 and 12]. It is worth noting that magnetic hysteresis could be observed even at 150 K in DFCO and this is attributed to the existence of short-range magnetic interactions above the long range ordering temperature of $T_1$. Such short ranged magnetic interactions are also proposed in Ref 19.

To reaffirm magnetic ordering behavior inferred above, heat-capacity was measured as a function of temperature from 2 to 150 K in the presence of different externally applied magnetic field for DFCO. To the best of our knowledge, this is for the first time that heat-capacity behavior is reported for the half-doped Dy orthochromite. For DFCO, a weak, but a distinct peak is observed around $T_2$, but such a peak around $T_1$ is not transparent in the raw data. In order to see the features more clearly, the temperature response of $dC/dT$, is plotted in the inset. The derivative curves show minima around $T_1 \sim 14$ K (left inset of Fig. 3a) and $T_2 \sim 121$ K (upper inset of Fig. 3a). These features in $C$ measurements confirm long range ordering. Additionally, the peak temperature around $T_2$ is seen to decrease even for a small application of the magnetic field, say, 2 and 5 kOe (right inset of Fig. 3a). Such a behaviour, that is, the decrease of the peak temperature for an application of magnetic-field, is characteristic of antiferromagnetic ordering. Figs. 3-(b) and 3(c) show the temperature dependence of $C$ for DEFCO and DGFCO. In both compounds a peak is observed around 121 K in $dC/dT$ (upper insets of Fig. 3-(b) and (c)), indicating that this transition is unaffected by rare-earth substitution and this is in agreement the inferences from susceptibility data. In the former compound, another minimum is observed around 10 K and this also matches with $T_2$ obtained from magnetization measurements. This peak temperature is shifted downwards with increasing field, as for the parent composition. For DGFCO compound, a mimima is observed around 3 K, confirming that $T_2$ is suppressed more dramatically for Gd substitution.

To investigate the effect rare-earth substitution on the magnetocaloric properties of DFCO, we studied the temperature dependence of the magnetocaloric effect (MCE) of this series. MCE is usually measured in terms of the change in isothermal magnetic entropy $|\Delta S_M|$ produced by changes in applied magnetic field. To obtain $|\Delta S_M|$ behaviour, $M(H)$ isotherms were acquired at different temperatures in the $T$ range of 2 to 150 K (in temperature interval of 2 K for 2 to 20 K and 114 to 130 K and 4 K for the rest). For this purpose, each isotherm was measured after cooling the sample from room temperature to the measurement temperature. $|\Delta S_M|$ was estimated by using the following expression [23]:

$$|\Delta S_M| = \sum [(M_n - M_{n+1})/(T_{n+1} - T_n)]\Delta H_n$$

where $M_n$ and $M_{n+1}$ are the magnetization values measured at field $H_n$ at temperatures $T_n$ and $T_{n+1}$ respectively. Figure 4 shows the $T$-dependent $|\Delta S_M|$ for different field variations from zero field. For DFCO, $\Delta S_M$ increases sharply and reaches a maximum value of ~ 10.8 J/kg-K around 10 K for $\Delta H$ = 50 kOe. This value of $\Delta S_M$ is slightly less than that observed in Ref. [19]. In the case of DEFCO, the value of $\Delta S_M$ increases to ~ 12 J/kg-K for the same field change. However, for DGFCO, a significant enhancement of $\Delta S_M$ is observed. The maximum value of $\Delta S_M$ for this compound is ~ 25 J/kg-K around 3.5 K. This value is less than that obtained for a single crystal of GdCrO$_3$ [11], but it is larger than those observed for other doped/undoped orthochromites [9, 12] and RMnO$_3$ multiferroics [24]. The observed change in $\Delta S_M$ is significant around $T_2$ only, but negligible around $T_1$. From the trends in MCE, we conclude that, while both Er and Gd enhance MCE below 10 K with respect to that for DFCO, a substitution of Gd results in a dramatic increase in the MCE values in these half-doped orthochromites.

We now address the dielectric behaviour of the materials. The samples are highly insulating which is an essential criterion for the dominance of intrinsic contributions to dielectric. This fact is also substantiated from the observed value of tanδ, which is very low. The $T$ responses of dielectric permittivity, both real ($\varepsilon'$) and loss part (tanδ) at 50 kHz of DFCO are shown in Fig. 5(a) and 5(b). The $\varepsilon'$ increases with increasing temperature; however, the increase is < 1.4% in the temperature range 2 to 150 K. An anomaly is observed around ~12.5 K. In order to look at the features more closely, the temperature derivative of $\varepsilon'$ is plotted in the insets of Fig. 5a and this derivate plot shows peaks at ~ 13 and 121 K, which are near the magnetic ordering temperature $T_2$ and $T_1$ respectively. Similarly, the temperature derivative of tanδ shows peaks at the same temperatures (insets of Fig. 5(b)). This is consistent with the presence of ME coupling in this compound. In this context it is to be noted that the observed anomalies are not frequency dependent. The features in dielectric constant is a consequence of magnetism due to cross-coupling between spin and lattice, however, the sharp increase in tanδ near 150 K could be due to decrease of its insulating behavior, therefore we have restricted our studies to temperatures below 150 K. Similar features are observed in the *T*-response of real and loss part of dielectric permittivity for DEFCO (Fig. 5(c)). For this compound, the dielectric anomalies are observed around 121 and 9 K, and these temperatures agree well with that obtained from magnetization measurements. In sharp contrast to the behavior of DFCO and DEFCO, for the material DGFCO, $\varepsilon'$ is seen to decrease with the increase in temperature up to ~20 K (Fig. 5(d)). Above this temperature, it increases as observed for the other two samples. As in the case of other two compounds, a dielectric anomaly is observed around 121 K. Hence, the persistence of the dielectric anomalies in the doped compounds indicates the presence of ME coupling in these mixed oxides as well.

The magnetic-field dependence of $\varepsilon'$ was measured at three temperatures: one below $T_2$ (3 K), between $T_2$ and $T_1$ (27 K) and above $T_1$ (125 K). The measurements were performed in the field range of 0 to 50 kOe and the data are shown for the frequency of 50 kHz in the form of $\Delta\varepsilon'$ where $\Delta\varepsilon' = [(\varepsilon'_H - \varepsilon'_{H=0})/\varepsilon'_{H=0}]$. In Fig. 6, panel (a) shows *H*-dependence of $\Delta\varepsilon'$ for DFCO at 3, 27 and 125 K, and panel (b) and (c) shows the same for DEFCO and DGFCO respectively at 3 and 27 K. For DFCO, the observed $\Delta\varepsilon'$ is clearly negative with a maximum magnitude of ~ 0.2% (say, at 3 K for 50 kOe). The coupling strength is greater than those observed for vanadates [25], but similar to that observed in other orthochromites [15]. The nature of the curve changes and the coupling strength decreases and reduces to zero with the increase of temperature across $T_2$. This feature indicates that a clear difference exists in the ME behaviour across the magnetic transitions. Interestingly, magnetic-field-induced effects in both *M*-(*H*) and $\Delta\varepsilon'$-(*H*) occur essentially near the same magnetic-field at the lowest measuring *T* (comparing insets in Fig. 6(a) and 2(e)). Additionally, it can be said that both magnetization (see Fig. 2(b)) and $\Delta\varepsilon'$ for a fixed magnetic field decrease as *T* increases consistent with ME coupling [26]. For DEFCO, at 3K, $\Delta\varepsilon'$ shows a positive peak followed by a sign reversal above ~ 35 kOe. $\Delta\varepsilon'$ also increases to 0.35%, which is 1.5 times to that observed for DFCO. At 27 K, the values get much smaller and the curve tends to be linear with a small positive value. Similar to DFCO, broad magnetic-field-induced effects occur in both *M*-(*H*) and $\Delta\varepsilon'$-(*H*) at the same field-range at the lowest measuring *T* (comparing inset of Fig. 6(b) and 2(e)). The magnitude of $\Delta\varepsilon'$ at 3 K increases significantly as a result of Gd substitution and is around 0.4% at 50 kOe.

To explore the presence of electric polarization in the magnetically ordered state, pyroelectric current measurements were performed. Following protocol was employed: The sample was cooled down to 7 K in the presence of poling electric field of ±250 kV/m. After that, the poling field was switched off and the sample was shorted to discharge any stray current. Pyroelectric current was measured during the warming cycle at a rate of 3 K/min. $P$ was obtained by integrating the pyroelectric current with respect to time. $T$-dependencies of $P$ are shown in Fig. 7. For DFCO, the maximum value of $P$ is obtained in the range $\sim 50 \mu C/m^2$, below magnetic ordering temperature $T_2$. For ferroelectricity, the orientation of electric domains should get changed by reversing the poling electric field. As expected, sign change of $P$ is observed by changing the direction of the poling field. This value of $P$ is less than the values observed in Ref. 19, but comparable to those observed for other orthochromites [15]. We now make some remarks about the origin of ferroelectricity below the magnetic ordering temperature. In $RCrO_3$, R-O-Cr magnetic interaction is responsible for ferroelectricity [14], whereas in $SmCrO_3$, it was attributed to a structural transition [27]. In order to explore possible structural anomalies in DFCO, we performed XRD measurements down to 24 K from room temperature. XRD pattern of DFCO confirmed that there is no structural transformation down to 24 K. The change in lattice parameter over this temperature range is < 0.2%. Therefore, the observed ferroelectricity is magnetic in its origin, as predicted earlier [19]. With the substitution of Gd/Er, the value of $P$ increases to about $\sim \pm 68 \ \mu C/m^2$ in DGFCO, and it further increases to $\sim \pm 100 \ \mu C/m^2$ in DEFCO. The enhanced value of $P$ in Gd/Er systems emphasizes that Gd/Er substitution may distort the lattice/adjust the bond lengths, thereby modifying the exchange interaction path Dy-O-Fe/Cr, which in turns enhance the ME coupling, polarization and MCE effect. Additionally, as observed from the figure, $P$ starts to develop above the magnetic ordering temperature of $T_1$ for all the compounds. It appears that short-range magnetic correlations persisting above $T_1$ can trigger the ferroelectricity, as demonstrated for $Fe_2TiO_5$, $Ca_3Co_2O_6$ and $Er_2BaNiO_5$ [28, 29].

Finally, it is important to note that the trend in MCE (see figure 4) by Gd/Er substitutions is similar to that of $P$ and ME coupling, as ME coupling is responsible for MCE, as demonstrated in Ref. 19. Similar proposal has been made in recent years in describing the theory of MCE in multiferroics materials [30].

**CONCLUSION**

We have investigated the effect of partial rare-earth (Er and Gd) substitution on the properties of half-doped orthochromite $DyFe_{0.5}Cr_{0.5}O_3$ using magnetic, dielectric and thermodynamic measurements. Our results reveal that the magnetic transition around 121 K is unaffected by rare-earth substitution, whereas the lower magnetic transition temperature is relatively suppressed. Multiferroic behavior persists in the Er and Gd doped systems. In all materials, dielectric anomalies are observed at the magnetic ordering temperatures. ME coupling is seen to get enhanced as a result of rare-earth substitution with respect to the parent compound. Additionally, both MCE and *P* are seen to get enhanced in the case of DEFCO and DGFCO. This enhancement is attributed to an increase in ME coupling in these compounds as compared to DFCO. These results establish that rare-earth substitution can be used to tune the ME coupling strength in these mixed metals oxides.

**Acknowledgments**

The authors acknowledge Kartik K. Iyer and V. Chandragiri for their help during experimental work. MKS and KM acknowledge experimental facilities of Advanced Material Research Centre (AMRC), IIT Mandi and also the financial support from IIT Mandi.

[#] Present Address: Experimental Physics V, Center for Electronic Correlations and Magnetism, University of Augsburg, D-86135 Augsburg, Germany

Table 1 Unit cell parameters obtained by Rietveld refinement for DFCO and its derivatives

| Parameters | DFCO | DEFCO | DGFCO |
|---|---|---|---|
| $a$ (Å) | 5.2860 (1) | 5.2624 (1) | 5.3133 (1) |
| $b$ (Å) | 5.5606 (1) | 5.5523 (1) | 5.5672 (0) |
| $c$ (Å) | 7.5902 (2) | 7.5719 (1) | 7.6192 (2) |
| $V$ (Å$^3$) | 223.02 (8) | 221.24 (7) | 225.38 (7) |
| Bragg R-factor | 3.60 | 4.51 | 5.98 |
| RF-Factor | 3.45 | 3.89 | 4.78 |
| $\chi^2$ | 1.880 | 2.24 | 2.14 |

**Figures**

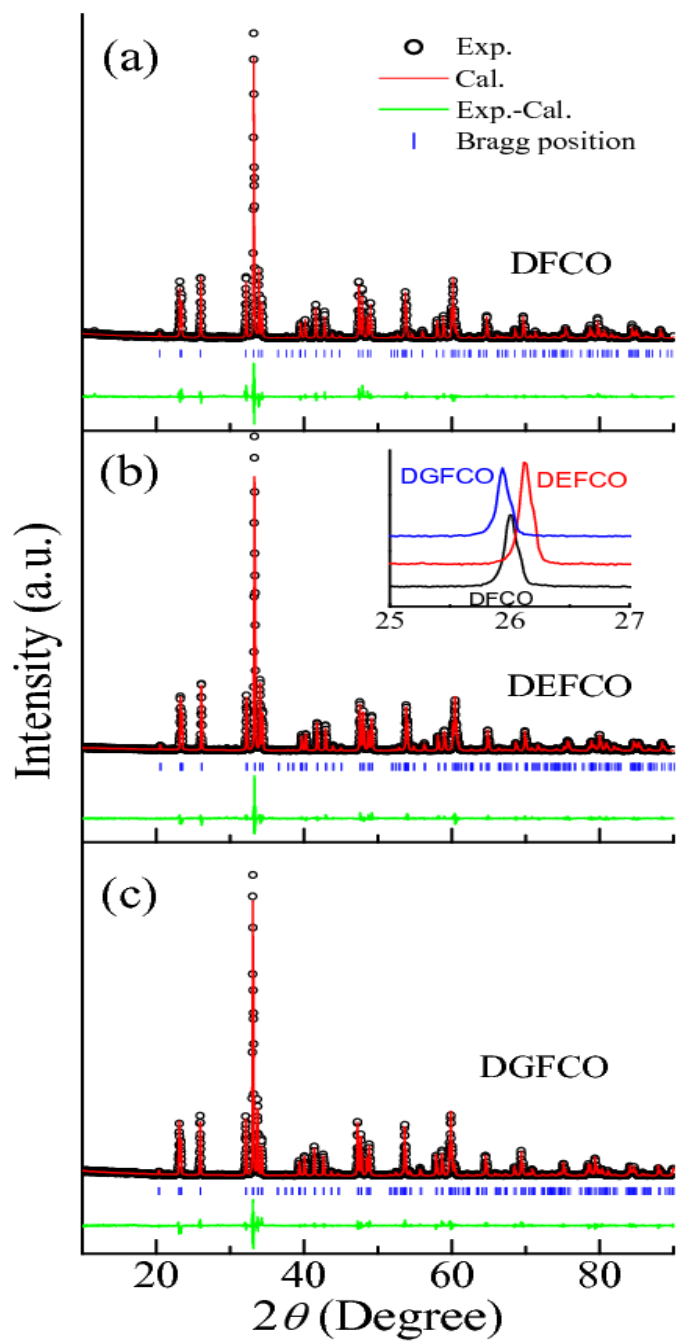

Figure 1. X-ray diffraction patterns (Cu K$_\alpha$) at room temperature: (a) Rietveld analysis results for DFCO. (b) Rietveld analysis results for DEFCO. Inset shows the pattern in an expanded form for one peak for all the samples, to bring out that the peaks shift with substitution. (c) Rietveld analysis results for DGFCO.

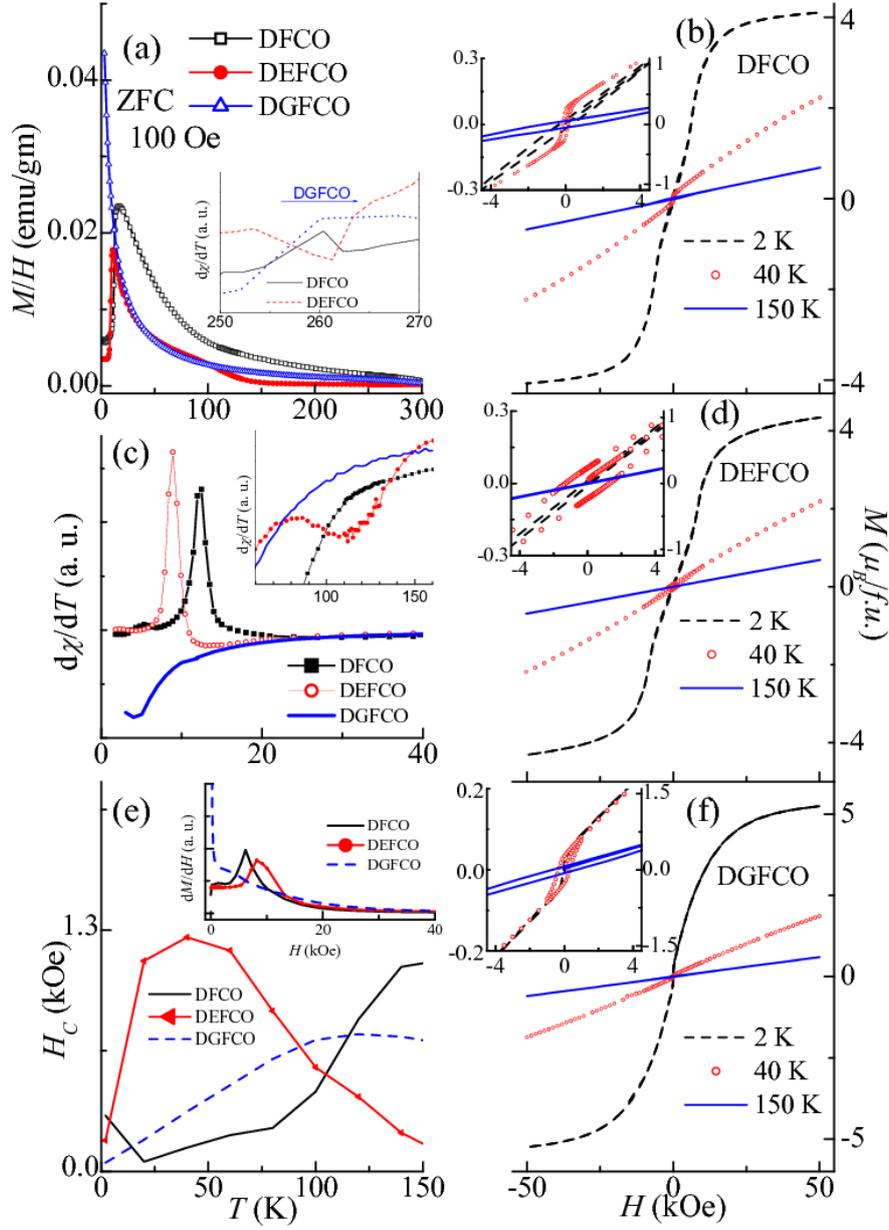

Figure 2. (a) Temperature ($T$) dependent dc susceptibility ($\chi = M/H$) curve obtained under zero-field-cooled condition with 100 Oe for the samples DFCO, DEFCO and DGFCO. Inset: $d\chi/dT$ plotted as a function of $T$ in the high $T$ region; (b) Isothermal magnetization curves for DFCO. Inset: $M(H)$ curves expanded in the low-field region. (c) $d\chi/dT$ plotted as a function of $T$ (in the low $T$ region) for these materials. Inset: The derivative plot in the intermediate $T$ region. (d) Isothermal $M(H)$ curve for DEFCO. Inset: $M(H)$ curves expanded at the low field region. (e) Coercive field vs. $T$ plots. Inset: $dM/dH$ plotted as a function of $H$ at 2 K. (f) Isothermal $M(H)$ curves for DGFCO. Inset: $M(H)$ curves expanded at the low field region. Unless stated, the lines through the data points serve as guides to the eyes.

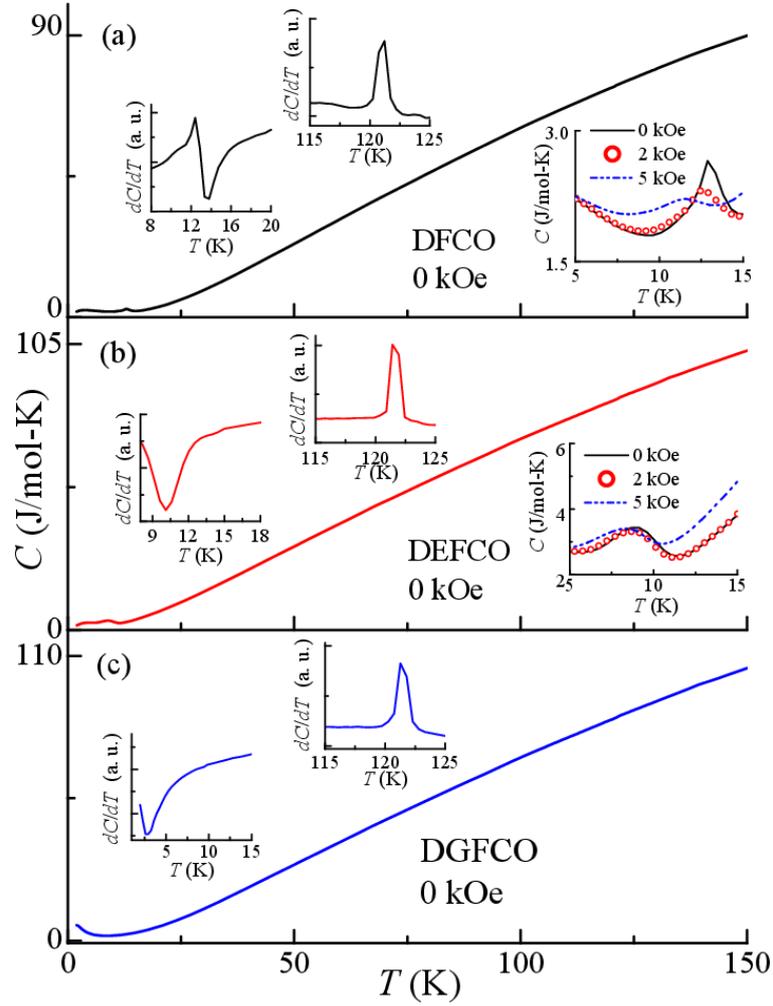

Figure 3. (a) Temperature (*T*) dependence of heat capacity (*C*) for DFCO at 0 kOe. Left inset: *dC/dT-T* plot in the *T* range of 8 to 20 K. Upper inset: *dC/dT-T* plot in the *T* range of 115 to 125 K. Right inset: Magnified *C* vs *T* plot (5 to 15 K) at different magnetic fields (*H*). (b) *T*-dependence of *C* for DEFCO at 0 kOe. Insets: Left inset: *dC/dT-T* plot in the *T* range of 8 to 18 K. Upper inset: *dC/dT-T* plot in the *T* range of 115 to 125 K for the same. Right Inset: Magnified *C* vs *T* plot (5 to 15 K) at different magnetic fields (*H*). (c) *T*-dependence of *C* for DGFCO at 0 kOe. Insets: Lower inset: *dC/dT-T* plot in the *T* range of 2 to 15 K. Upper inset: *dC/dT-T* plot in the *T* range of 115 to 125 K.

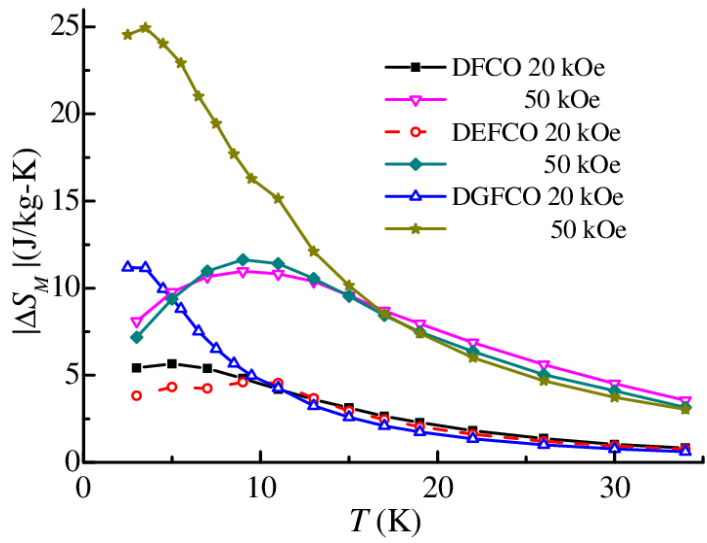

Figure 4. Isothermal magnetic entropy change plotted as a function of temperature for the three materials, DFCO, DEFCO and DGFCO for a variation of the magnetic field from 0 to 20 and 50 kOe.

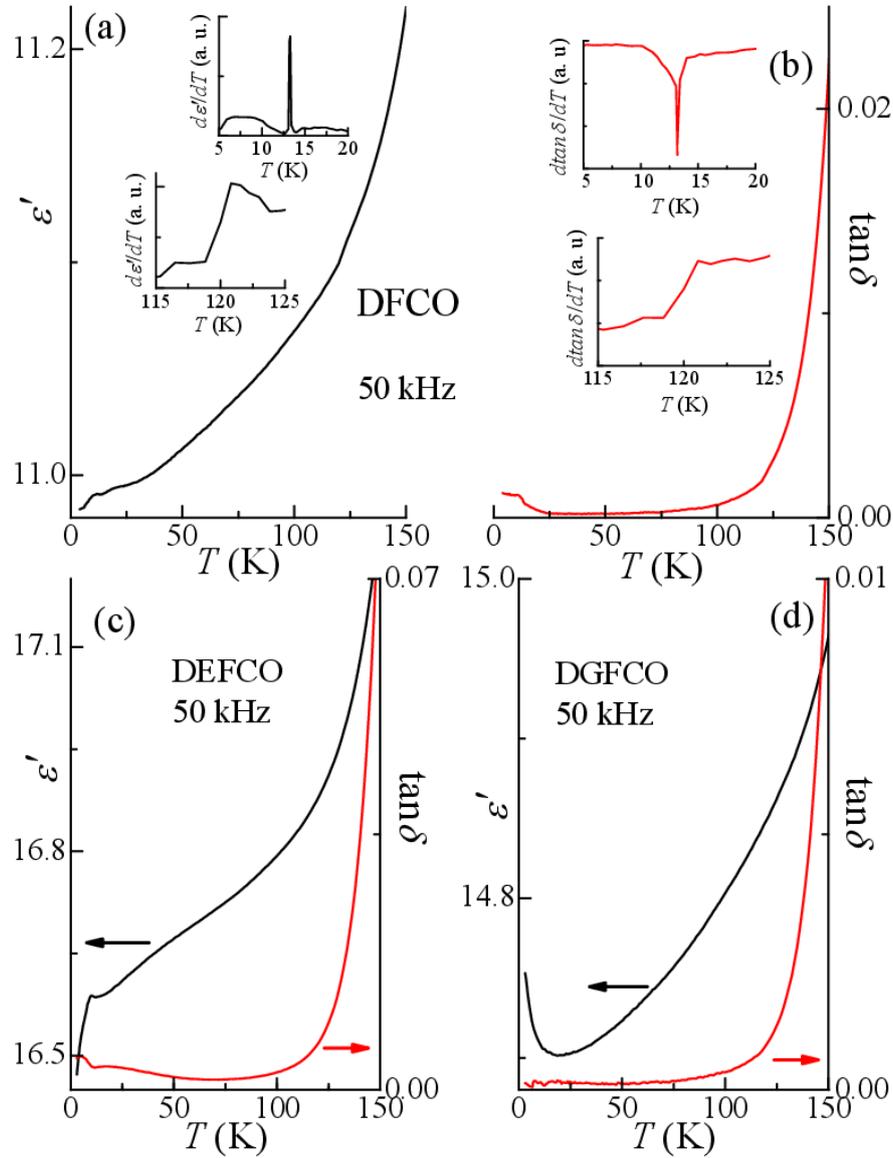

Figure 5. (a) Temperature ($T$) dependence of real part of dielectric permittivity ($\varepsilon'$) for DFCO measured with 50 kHz. Inset: Upper inset shows the $d\varepsilon'/dT$-$T$ plot in the $T$ range of 5 to 20 K. Lower inset shows the $d\varepsilon'/dT$-$T$ plot in the $T$ range of 115 to 125 K. (b) $T$-dependent dielectric loss (tan$\delta$-$T$) plot for DFCO measured with 50 kHz. Inset: Upper inset shows the $d(\tan\delta)/dT$-$T$ plot in the $T$ range of 5 to 20 K. Lower inset shows the $d(\tan\delta)/dT$-$T$ plot in the $T$ range of 115 to 125 K. (c) $T$-dependence of $\varepsilon'$ (left axis) and tan$\delta$ (right axis) for DEFCO at 50 kHz. (d) $T$-dependence of $\varepsilon'$ (left axis) and tan$\delta$ (right axis) for DGFCO at 50 kHz. The measurements were carried in the absence of external magnetic field.

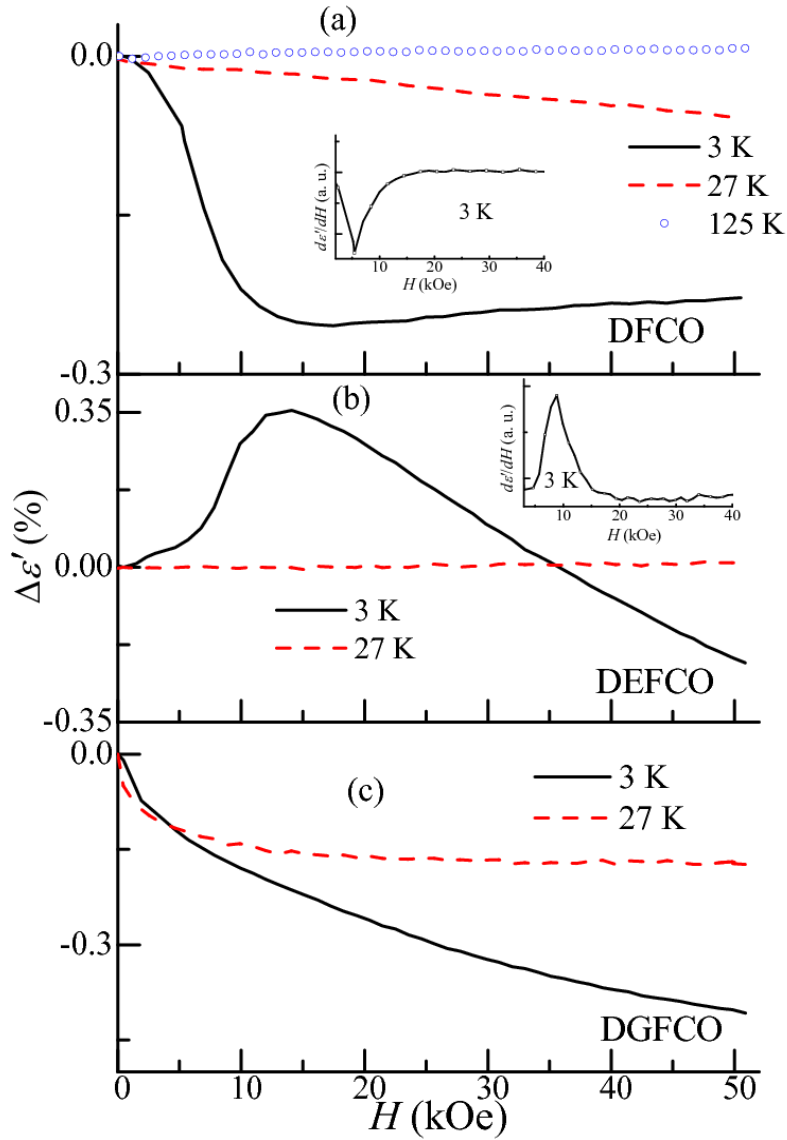

Figure 6. Relative change in dielectric constant by the application of external magnetic field for DFCO, DEFCO and DGFCO at different temperatures ($T$). (a) $\Delta\varepsilon'$ vs. $H$ plot for DFCO. Inset: $d\varepsilon'/dH$ vs. $H$ for the same at 3 K. (b) $\Delta\varepsilon'$ vs. $H$ plot for DEFCO. Inset: $d\varepsilon'/dH$ vs. $H$ plots for the same at 3 K. (c) $\Delta\varepsilon'$ vs. $H$ plot for DGFCO.

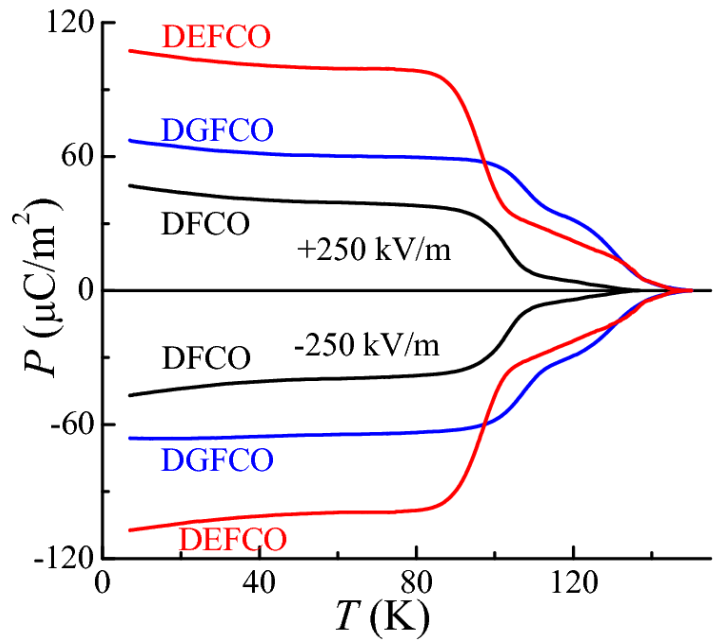

Figure 7. Temperature dependence of electric polarization for the materials, DFCO, DEFCO, and DGFCO.